\documentclass[aps,pre,twocolumn,floatfix,longbibliography,superscriptaddress]{revtex4-1}
\pdfoutput=1
\usepackage[normalem]{ulem}
\usepackage{float}
\usepackage{color}
\usepackage{graphicx}
\usepackage{amsfonts}
\usepackage{amsmath}
\usepackage{mathtools}
\usepackage{bbold}
 \usepackage{pdfpages}
\usepackage{pgffor}
\makeatletter
\AtBeginDocument{\let\LS@rot\@undefined}
\makeatother
\usepackage[colorlinks]{hyperref}
\hypersetup{
    colorlinks,
    linkcolor={blue},
    citecolor={blue},
    urlcolor={blue}
}
\usepackage{url}

\newcommand{\dd}{\mathrm{d}}

\newcommand{\av}[1]{\left\langle #1 \right\rangle}

\newcommand{\Eq}[1]{Eq.~\ref{#1}}

\newcommand{\Sdot}{\dot{S}}

\newcommand{\la}{\langle}
\newcommand{\ra}{\rangle}

\newcommand{\x}{\mathbf{x}}
\newcommand{\y}{\mathbf{y}}

\newcommand{\I}{\mathcal{I}}

\newcommand{\mcal}{\mathcal}
\newcommand{\cvec}{\mathbf{c}}
\newcommand{\sigmaF}{\sigma^2_{\widehat{\mathcal{F}}}}
\newcommand\chout{\bgroup\markoverwith{\textcolor{red}{\rule[0.5ex]{2pt}{1.0pt}}}\ULon}
\newcommand\fedout{\bgroup\markoverwith{\textcolor{blue}{\rule[0.5ex]{2pt}{1.0pt}}}\ULon}
\newcommand\grzout{\bgroup\markoverwith{\textcolor{magenta}{\rule[0.5ex]{2pt}{1.0pt}}}\ULon}

\DeclareMathOperator{\Tr}{\rm Tr}

\newcommand{\p}{\mathbf{pc}}
\newcommand{\Fb}{\mathbf{F}}
\newcommand{\Db}{\mathbf{D}}

\newcommand{\Fex}{\mathcal{F}_{\rm ex}}
\newcommand{\Fhat}{\widehat{\mathcal{F}}}

\begin{document}

\title{Learning the Non-Equilibrium Dynamics of Brownian Movies}

\author{Federico S. Gnesotto}
 \author{Grzegorz Gradziuk}
\affiliation{Arnold-Sommerfeld-Center for Theoretical Physics and Center for
  NanoScience, Ludwig-Maximilians-Universit\"at M\"unchen,
   D-80333 M\"unchen, Germany.}
   
\author{Pierre Ronceray}
\email{ronceray@princeton.edu}
\affiliation{Center for the Physics of Biological Function, Princeton University, Princeton, NJ 08544, USA}

\author{Chase P. Broedersz}
\email{c.broedersz@lmu.de}
\affiliation{Arnold-Sommerfeld-Center for Theoretical Physics and Center for
  NanoScience, Ludwig-Maximilians-Universit\"at M\"unchen,
   D-80333 M\"unchen, Germany.}

\date{\today}

\begin{abstract}
Time-lapse microscopy imaging provides direct  access to the dynamics of soft and  living systems. At mesoscopic scales, such microscopy experiments reveal intrinsic fluctuations, which may have both thermal and non-equilibrium origins. These intrinsic fluctuations, together with measurement noise, pose a major challenge for the  analysis of the dynamics of these ``Brownian movies". Traditionally, methods to analyze such experimental data rely on tracking embedded or endogenous probes. However, it is in general unclear how to select appropriate tracers; it is not evident, especially in complex many-body systems, which degrees of freedom are the most informative about their non-equilibrium nature. Here, we introduce an alternative, tracking-free approach that overcomes these difficulties via an unsupervised analysis of the Brownian movie. We develop a dimensional reduction scheme that selects a basis of modes based on dissipation, and we subsequently learn the non-equilibrium dynamics in this basis and estimate the entropy production rate. In addition, we  infer time-resolved force maps in the system and show that this approach is scalable to large systems, thus providing a potential alternative to microscopic force-probes. After benchmarking our method against a minimal two-beads model, we illustrate its broader applicability with an example inspired by active biopolymer gels.

\end{abstract}
\maketitle
\noindent 

\begin{figure*}
\centering
\includegraphics[width=\textwidth]{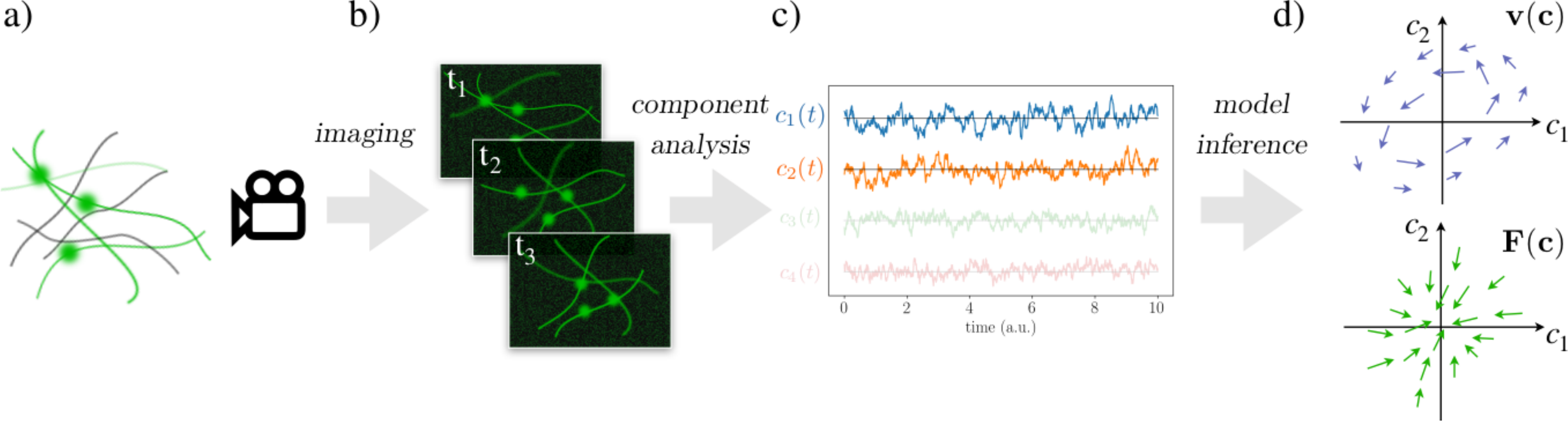}
\caption{\textbf{Schematic illustration of our approach to learn non-equilibrium dynamics from a Brownian movie}. a) Sketch of a network of biopolymers (black) with embedded fluorescent filaments and beads (green). b) Image-frames of the fluorescent components in panel a) at three successive time points. c) The time trajectories of the projection coefficients $c_1(t),c_2(t), \cdots$: the coefficients and respective trajectories discarded by the dimensional reduction are faded. d) Sketch of the the inferred velocity $\mathbf{v}(\cvec)$ (top) and force field $\Fb(\cvec)$ (bottom) in the space $\{c_1,c_2\}$.}
\label{Fig1}
\end{figure*}

\noindent Over the last two centuries, fundamental insights have been gleaned about the physical properties of biological and soft
matter systems by using microscopes to image their dynamics~\cite{stephens_light_2003,sahl_fluorescence_2017}. 
At the micrometer scale and below, however, this dynamics is inherently stochastic, as ever-present thermally driven Brownian fluctuations give rise to 
short-time displacements~\cite{brown_xxvii._1828, einstein_uber_1905,frey_brownian_2005}. This random motion makes such ``Brownian
movies'' appear jiggly and erratic; this randomness is further exacerbated by measurement noise and
limited resolution intrinsic to, \textit{e.g.}, fluorescence microscopy~\cite{waters_accuracy_2009}. In light of all these
sources of uncertainty, how can one best make use of measured Brownian movies of a systems dynamics, to learn the underlying physics of the fluctuating and persistent forces? 

In addition to thermal effects, active processes can strongly impact the stochastic dynamics of a system~\cite{mackintosh_active_2010,gnesotto_broken_2018,aranson_active_2013,cates_motility-induced_2015,fodor_how_2016}. Recently, there has been a  growing interest in quantifying and characterizing the
non-equilibrium nature of the stochastic dynamics in active soft and living systems~\cite{martinez_inferring_2019, guo_probing_2014,fakhri_high-resolution_2014,turlier_equilibrium_2016,battle_broken_2016,gladrow_broken_2016,mura_nonequilibrium_2018, seara_entropy_2018,ma_active_2014,li_quantifying_2019,sanchez_spontaneous_2012, frishman_learning_2019, roldan_estimating_2010}.
In cells,  molecular-scale activity, powered for instance by ATP hydrolysis,
control mesoscale non-equilibrium processes in assemblies such as cilia ~\cite{sanchez_cilia-like_2011,battle_intracellular_2015}, flagella ~\cite{riedelkruse_how_2007}, chromosomes \cite{weber_nonthermal_2012}, protein droplets ~\cite{brangwynne_active_2011} or cytoskeletal networks ~\cite{mizuno_nonequilibrium_2007,brangwynne_cytoplasmic_2008,koenderink_active_2009,brangwynne_nonequilibrium_2008}. The irreversible nature of such non-equilibrium processes can lead to measurable dissipative currents in a phase space of mesoscopic degrees of freedom~\cite{battle_broken_2016,gladrow_broken_2016,gnesotto_broken_2018, paijmans_discrete_2016,kimmel_inferring_2018,wan_time_2018}. Such dissipative currents can be quantified by the entropy production rate~\cite{seifert_stochastic_2012, seara_entropy_2018, li_quantifying_2019, mura_nonequilibrium_2018, frishman_learning_2019}, but it remains an outstanding challenge to accurately infer the entropy production rate by analyzing Brownian movies of such systems. 

Traditional approaches to measure microscopic forces and analyze time-lapse microscopy data typically rely on tracking the position or shape of
well-defined probes such as tracer beads, fluorescent proteins and filaments, or simply on exploiting the natural
contrast of the intracellular medium to obtain such tracks~\cite{mizuno_nonequilibrium_2007,turlier_equilibrium_2016, crocker_methods_1996,battle_broken_2016, weber_nonthermal_2012,guo_probing_2014, fakhri_high-resolution_2014,brangwynne_nonequilibrium_2008,levine_one-_2000,sawada_force_2006, grashoff_measuring_2010}. The tracer trajectories 
can be studied through stochastic analysis techniques to
extract an effective model for their dynamics and infer quantities like the entropy
production~\cite{mura_nonequilibrium_2018, mura_mesoscopic_2019,frishman_learning_2019,seara_entropy_2018,bruckner_stochastic_2019,selmeczi_cell_2005,stephens_dimensionality_2008,li_quantifying_2019}.  There are, however, many cases in which tracking is
impractical~\cite{seara_dissipative_2019,edera_differential_2017}, due to limited resolution or simply because there are no
recognizable objects to use as tracers. Another, more fundamental limitation of tracking
is that one then mostly learns about the dynamics of the tracked object---not of the system as a whole. Indeed, the dissipative power in a system might
not couple directly to the tracked variables, and \textit{a priori}, it might not be clear which coordinates will be most informative about such dissipation. This raises the question how one can identify
which degrees of freedom best encode the forces and non-equilibrium dissipation in a given system.

Here we propose an alternative to tracking: learning the dynamics and
inferring the entropy production directly from the unsupervised
analysis of Brownian movies. We first decompose the movie into generic
principal modes of motion, and predict which ones are the most likely
to encode useful information through a ``Dissipative Component
Analysis'' (DCA). We then perform a dimensional reduction, which leads
to a representation of the movie as a stochastic trajectory in this
component space. Finally, we employ a recently introduced method,
Stochastic Force Inference (SFI)~\cite{frishman_learning_2019}, to
analyze such trajectories. Our approach not only yields an estimate of
the entropy production rate of a Brownian movie, which is a controlled
lower bound to the system's total entropy production, but also
important dynamical information such as a time-resolved force map of
the imaged system. Thus, our approach may provide an alternative to methods that use microcopic force sensors ~\cite{sawada_force_2006, grashoff_measuring_2010, lucio_chapter_2015, han_cell_2018}.
 In this article, we first present the method in its
generality, then benchmark it on a simple two-beads model. Finally, we demonstrate the potential of our approach on simulated semi-realistic
fluorescence microscopy movies of out-of-equilibrium biopolymer
networks, and we show that the force inference approach is scalable to large systems. 

\section{Principle of the method}\label{sec:principle}

\noindent We begin by  describing a tracking-free method to infer
 the dynamical equations of a system from raw image
sequences. This approach allows us to determine a bound on the dissipation of a system, as well as the force-field in image space. 

Our starting point is the assumption that the physical system we
observe (Fig.~\ref{Fig1}a)---such as a cytoskeletal network or a fluctuating
membrane---can be described by a configurational state vector $\mathbf{x}(t)$ at time $t$, undergoing
steady-state Brownian dynamics in an unspecified $d$-dimensional
phase space:
\begin{equation}
  \label{eq:Lang_initial} 
  \frac{\dd \mathbf{x}}{\dd t} = \mathbf{\Phi}(\mathbf{x})  + \sqrt{2\mathbf{D}(\mathbf{x})} \xi(t),
\end{equation}
where $\mathbf{\Phi}(\mathbf{x})$ is the drift field,
$\mathbf{D}(\mathbf{x})$ is the diffusion tensor field, and throughout
this article $\xi(t)$ is a Gaussian white noise
vector 
($\la \xi(t) \ra=0$ and
$\la \xi_i(t) \xi_j (s) \ra =\delta_{ij}\delta(t-s) $). Note that when
diffusion is state-dependent, $\sqrt{2\mathbf{D}(\mathbf{x})} \xi(t)$
is a multiplicative noise term: we employ the It\^o convention for the
drift, \emph{i.e.} $\mathbf{\Phi}(\mathbf{x})= \mathbf{F(\mathbf{x})}+\nabla \cdot \mathbf{D}(\mathbf{x})$, where $\mathbf{F(\mathbf{x})}$ is the physical  force in the absence of  Brownian noise ~\cite{lau_state-dependent_2007,risken_fokker-planck_1996}.

Our goal is to learn as much as possible about the process described by \Eq{eq:Lang_initial} from an experimental observation. In particular, we aim to measure if, and how far, the system is out-of-equilibrium by determining the irreversible nature of its dynamics. This irreversibility is quantified by the system's entropy production rate~\cite{seifert_stochastic_2012}
\begin{equation}
  \label{eq:Sdot}
  \Sdot_\mathrm{total} = \av{ \mathbf{v}(\mathbf{x}) \mathbf{D}^{-1}(\mathbf{x}) \mathbf{v}(\mathbf{x})  },
\end{equation}
where $\av{ \cdot }$ denotes a steady-state average, throughout this article we set Boltzmann's constant
$k_{\rm B} =1$, and 
$\mathbf{v}(\mathbf{x})$ is the mean phase space velocity field quantifying
the presence of irreversible currents. Specifically, using the steady-state Fokker-Planck
equation one can write
$\mathbf{v}(\mathbf{x}) = \mathbf{F}(\mathbf{x}) - \mathbf{D}(\mathbf{x}) \nabla \log P(\mathbf{x})$,
where $P(\mathbf{x})$ is the steady-state probability density function, and flux balance imposes that $\nabla \cdot (P \mathbf{v}) =0$.

The input of our method consists  of a discrete time-series of
microscopy images of the physical system $\{\I(t_0), \dots \I(t_N)\}$---a ``Brownian movie'' (Fig.~\ref{Fig1}b). Each image $\I(t)$ is an imperfect
representation of the state $\mathbf{x}(t)$ of the physical system as
a bitmap, \emph{i.e.} a $L\times W$ vector of real-valued pixel
intensities~\footnote{We neglect the discretization effect induced by
  the finite number of pixel intensities here.}. Specifically, we
model the imaging apparatus as a noisy nonlinear map $\I(t) = \bar{\I}(\mathbf{x}(t)) + \mathcal{N}(t)$, where $\mathcal{N}$
is a temporally uncorrelated noise representing measurement noise
(such as the fluctuations in registered fluorescence intensities), and
$\bar{\I}(\mathbf{x})$ is the ``ideal image'' returned on average by
the microscope when the system's state is $\mathbf{x}$. We assume that
this map $\bar{\I}(\mathbf{x})$ is time-independent (\textit{i.e.} that
the microscope settings are fixed and stable).

Importantly, if no information is lost by the imaging process, the ideal image $\bar{\I}(t)$ undergoes a Brownian dynamics equation determined by the nonlinear transformation of \Eq{eq:Lang_initial} through the map $\mathbf{x}\mapsto\bar{\I}(\mathbf{x})$, as prescribed by 
It\^o's lemma~\cite{oksendal_stochastic_2003}. 
In general, however, there is information loss and $\bar{\I}(\mathbf{x})$ is not invertible: due to finite resolution or because some elements are simply not visible, the imaging may not capture the full high-dimensional state of the system.  For
this reason, the dynamics in image space are not uniquely
specified by the ideal image value $\bar{\I}$; they also depend on
``hidden'' degrees of freedom $\x_h$ not captured by the image. In this case, a
Markovian dynamical equation for $\bar{\I}$ alone does not exist, but by  including the dynamics of $\x_h$, we can write 
\begin{equation}
\frac{\dd}{\dd t}(\bar{\I},\x_h) =  \phi(\bar{\I},\x_h) + \sqrt{2\mcal D(\bar{\I},\x_h) }\xi(t),
\label{eq:Langevin_images_new}
\end{equation}
Here $\phi(\bar{\I},\x_h)$ and $\mcal D(\bar{\I},\x_h)$ are  the drift field and diffusion tensor, respectively, in the combined space of pixel intensities and hidden variables. Our Brownian movie analysis allows us to infer the mean image drift $\phi(\bar{\I})\coloneqq\av{\mcal \phi_{\bar{\I}}(\bar{\I},\x_h)|\bar{\I}}$ and mean image diffusion tensor $\mcal D(\bar{\I})\coloneqq\av{\mcal D_{\bar{\I}}(\bar{\I},\x_h)|\bar{\I}}$, averaged over the degrees of freedom $\x_h$ lost in the imaging process. From drift and diffusion fields we can directly obtain the mean image force field $\mcal F(\bar{\I})=\phi(\bar{\I})-\nabla \cdot \mcal D(\bar{\I})$. Similar to force and diffusion fields, the phase space currents
$\mathbf{v}(\mathbf{x})$ in the $d$-dimensional physical phase space,
transform into currents $\mathcal{V}(\bar{\I})$ in the
$L\times W$-dimensional image space---again, averaged over
unobserved degrees of freedom. These currents result in an apparent
entropy production associated to the image dynamics~\footnote{Note that
  we consider here only the entropy production associated to apparent
  currents. The irreversible dynamics of unobserved degrees of freedom
  has repercussion on non-Markovian effects in the dynamics, which
  result in other contributions to the entropy 
 production~\cite{roldan_estimating_2010}, which we
  neglect here.},
\begin{equation}
  \label{eq:Sdot_apparent}
  \Sdot_\mathrm{apparent} = \av{ \mathcal{V}(\bar{\I}) \mathcal{D}^{-1}(\bar{\I}) \mathcal{V}(\bar{\I}) }.
\end{equation}
Importantly, the function
$(\mathcal{V},\mathcal{D})\mapsto
\mathcal{V}\mathcal{D}^{-1}\mathcal{V}$ is multivariate convex, and
thus by Jensen's inequality,
$\Sdot_\mathrm{apparent} \leq \Sdot_\mathrm{total}$: the apparent
entropy production is a lower bound to the total entropy production.

The goal of our method is to reconstruct the mean image-space dynamics
$(\mcal F(\bar{\I}),\mcal D(\bar{\I}))$, and in particular the
corresponding entropy production (\Eq{eq:Sdot_apparent}). However, doing so in the high-dimensional image
space is unpractical and would require unrealistic amounts of data. We
therefore need to reduce the dimensionality of our system to a
tractable number of relevant degrees of freedom.

Because each image represents a physical state of the system, we expect that the ideal images $\bar{\I}(t)$ all share similar structural features. Consequently, the Brownian movie occupies only a smaller subspace in the space of all configurations of pixel intensities. To restrict ourselves to the manifold of images representing the physical states, we  can either perform only Principal
Component Analysis (PCA) or, as we shall see later, reinforce PCA with an  analysis which provides an additional basis transformation to select the most dissipative components. The
idea behind this  approach is that the components are hierarchically ordered according to how much they contribute to the entropy production, such that it becomes possible to  truncate the basis and reduce the dimensionality of the problem, while retaining maximum information about the system's irreversibility.

We truncate the basis of components according to three criteria: 1) Noise floor---due
  to the finite amount of data and the measurement noise present in
  the Brownian movie, some modes are indistinguishable from the measurement
  noise. We only keep modes that rise above this noise floor.  2) Time resolution of the dynamics---we only
  consider the components whose statistical properties are consistent
  with Brownian dynamics, \textit{i.e.} such that the short-time
  diffusive behavior can be resolved through the noise. 
  3) Dimension of phase space---for a physical system $\x(t)$ with $d$ observable degrees of freedom the ideal images $\bar{\I}(\mathbf{x}(t))$ will form a $d$-dimensional manifold in the large $(L\times W)$-dimensional image space. Depending on the shape of the manifold it may be possible to project the images $\I(t)$ on an appropriate $d$-dimensional linear subspace, without losing any information about the dynamics of $\x(t)$. This restriction ensures that the  dynamics is inferred in a space of dimensionality smaller or equal to that of the physical system, thus avoiding singularities in the inference of diffusion and related quantities. We determine the dimensionality of the manifold $d$ by performing PCA locally, in a region where the manifold is approximately flat and keep only the first $d$ principal components of the globally performed PCA (see Supplementary Material Sec. V).
  Note, while these first $d$ components will be sufficient to represent the dynamics of $\x(t)$, more modes may be need to faithfully reconstruct configurational and dynamical quantities in image space.

Our task is now reduced to inferring the mean dynamics in  component space,
\begin{equation}
\mathbf{\Phi}(\cvec) \coloneqq \av{ \mathbf{\Phi}_{\cvec}(\cvec,\x_h)|\cvec} \ ,\  
\mathbf{D(\cvec)} \coloneqq\av{\Db_{\cvec}(\cvec,\x_h)|\cvec}
\label{eq:Langevin_components}
\end{equation}
where $\cvec(t)=(c_1(t),c_2(t),\cdots,c_n(t))$ are the components obtained after a linear transformation of the images (see Fig.~\ref{Fig1}c), and the hidden degrees of freedom $\x_h$ now also include those present in the image, but left out after the components' truncation. This procedure has reduced the system's dynamics to that of a smaller
number of components, making it possible to learn $\mathbf{\Phi}(\cvec)$ and $\Db(\cvec)$. 

To this end, we employ a recently introduced method, Stochastic Force Inference~\cite{frishman_learning_2019}
(SFI), for the inverse Brownian dynamics problem. Briefly, this
procedure is based on a least-squares approximation of the diffusion
and drift fields using a basis of known functions (such as
polynomials). This method is data-efficient, not limited to
low-dimensional signals or equilibrium systems, robust against
measurement noise, and  provides estimates of the inference
error, making it well suited for our purpose. In practice, we use SFI in two ways: 1) we infer the velocity
field $\mathbf{v}(\mathbf{c})$ (Fig.~\ref{Fig1}d) and the diffusion field
$\mathbf{D}(\mathbf{c})$, which we use to measure
the entropy production. 2) We
 infer the drift field $\mathbf{\Phi}(\mathbf{c})$, compute the image force $\Fb(\mathbf{c})=\mathbf{\Phi}(\mathbf{c})-\nabla \cdot \Db(\cvec)$  (Fig.~\ref{Fig1}d), and thus reconstruct the
dynamics of the components. To render this deterministic dynamics more intelligible,
we can transform $\Fb(\mathbf{c})$ back into image space by inverting the
$\mathcal{I}\mapsto \mathbf{c}$ linear transformation: this results in
a ``pixel force'' map, which indicates at each time step
which pixel intensities tend to increase or decrease.  This provides,
we argue, a novel way to gain insight into the dynamics of Brownian
systems and disentangle deterministic forces from Brownian motion
without tracking.

Our analysis framework can thus be schematically summarized as:
\textit{imaging $\to$ component analysis $\to$ model inference} (Fig.~\ref{Fig1}). This procedure
allows the inference of entropy production and reconstruction of the
dynamical equations from image sequences of a Brownian system.

\section{A minimal example: two-beads Brownian movies}

\begin{figure*}
\centering
\includegraphics[width=\textwidth]{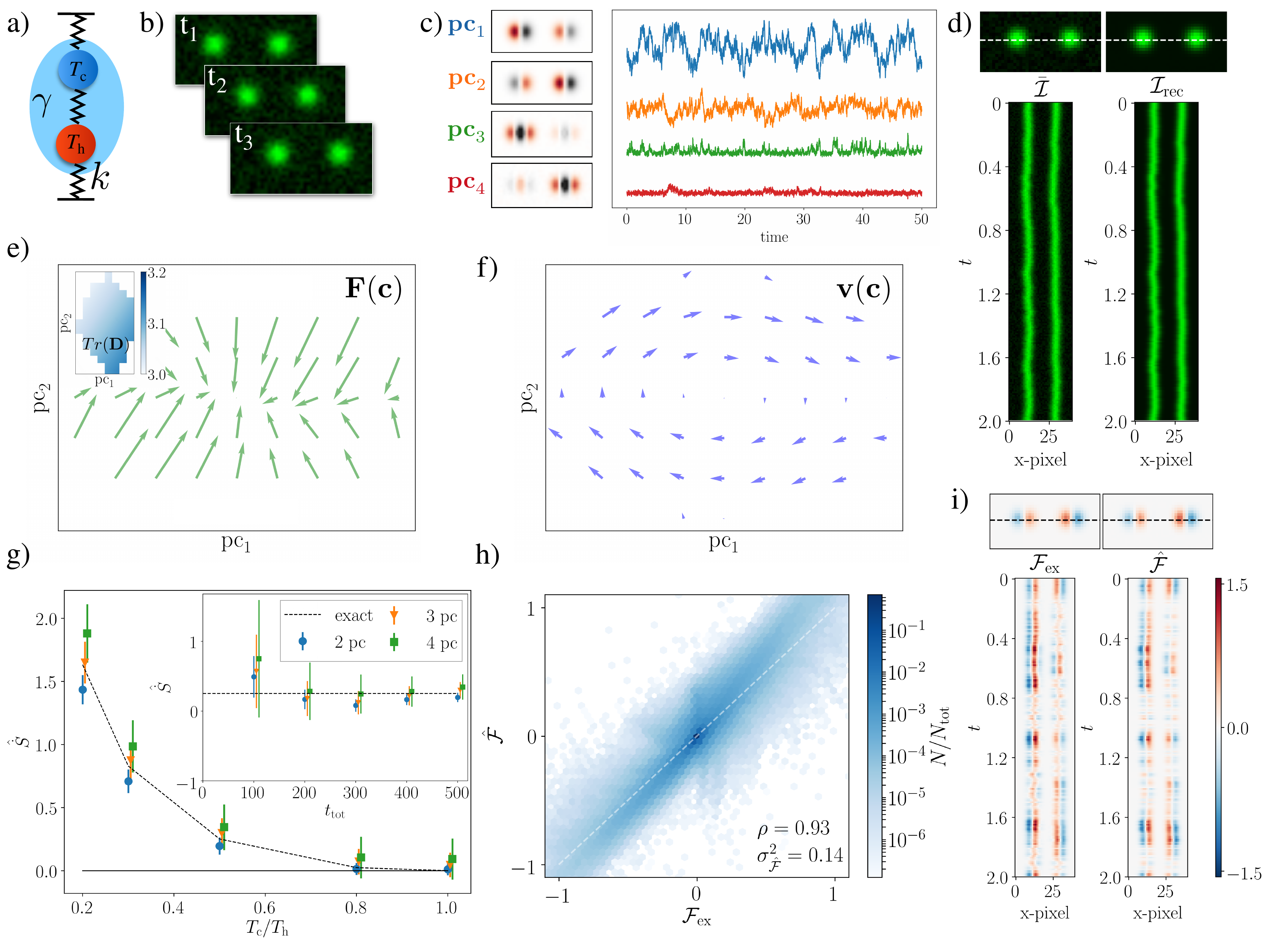}
\caption{\textbf{Benchmarking the Brownian movie learning approach with a simple toy model} a) Schematic of the two-bead model. We use $k=2$, $\gamma = 1$, $k_B=1$; the temperature of the hot bead $T_{\rm h}=1$ is fixed and the temperature of the cold bead $T_{\rm c} \leq 1$ is varied. b)  $40\times20$  Frames of the noisy ($10 \%$ noise) Brownian movie for the two bead-model at successive time-points c) The first 4 principal components with  time-traces of respective projection coefficients. The color map displays negative pixel values in black and positive pixel values in red. d) Top: Snapshot of the exact image $\I_{\rm ex}$ (left) and the reconstructed image $\I_{\rm rec}$ (right) reconstructed with the first four principal components. Bottom: associated kymographs. We compare pixel intensities along the superimposed horizontal dashed line. e) Force field in the space of the first two principal components $\p_1 \times \p_2 $. Inset:  trace of diffusion tensor $\Tr(\Db)$. f) The mean phase space velocity in  $ \p_1 \times \p_2  $.  g) Inferred entropy production rate $\widehat{\dot{S}}$ for varying  temperature ratio $T_{\rm c}/T_{\rm h}$ and number of included principal components. Inset: $\widehat{\dot{S}}$ as a function of trajectory length for a fixed $T_{\rm c}/T_{\rm h}=0.5$. h) Scatter plot of the exact image force field $\mathcal{F}_{\rm ex}$  vs. the inferred image force field $\hat{\mathcal{F}}$ for different pixels and time points (data has been binned for visualization purposes). Results are obtained using the first four principal components. i) Top: comparison of inferred $\hat{ \mathcal{F}}$ and exact $\mathcal{F}_{\rm ex}$ image-space force fields. Bottom: associated kymographs. Panels c)-d)-e)-f)-h)-i) have been obtained with $T_{\rm c}/T_{\rm h}=0.5$ and for a trajectory of length $t_{\rm tot}=10^5 \Delta t$, $\Delta t=0.01$. Panel g) with $t_{\rm tot}=5\times10^4 \Delta t$.  The SFI routine was employed with a first order polynomial basis for the inference of forces and diffusion fields. The  noise-corrected estimator was used to infer the diffusion fields~\cite{frishman_learning_2019}.}
\label{Fig2}
\end{figure*}  

Next, we test the performance of our procedure on a simple
non-equilibrium model: two coupled beads moving in one
dimension. The beads are coupled by Hookean springs with stiffness $k$ and experience Stokes drag with friction coefficient $\gamma$, due to the surrounding fluid 
(Fig.~\ref{Fig2}a).  In this two-bead model, the time-evolution of the
bead displacements $\x(t)=(x_1(t),x_2(t))$ obeys the overdamped
Langevin Eq. \eqref{eq:Lang_initial}, with $\mathbf{F}(\x)=\mathbf{A} \x$ and
$A_{ij} = (1- 3\delta_{ij}) k/\gamma$. The system is driven out of
thermodynamic equilibrium by imposing different temperatures on the
two beads: $D_{ij}= \delta_{i j} k_{\rm B}T_i / \gamma $~\cite{crisanti_nonequilibrium_2012,berut_theoretical_2016, li_quantifying_2019, gnesotto_broken_2018,gnesotto_nonequilibrium_2019}.
 First, we obtain position trajectories for the two beads by
discretizing their stochastic dynamics using an Euler integration
scheme (see Supplementary Material Sec. I). Then,  we use these position trajectories to 
construct a noisy Brownian movie (Fig.~\ref{Fig2}b) (cf. Supplementary Material Sec. II and Supplementary Movie 1). Note that by construction, the
steady-state dynamics of the two-beads system in image space is
governed by a non-linear Langevin equation with multiplicative noise.

We seek to reduce the dimensionality of the data by finding
relevant components. To this end, we employ Principal Component
Analysis (PCA)~\cite{bishop_pattern_2006} and determine the basis of $n$ principal components
$\p_1, \p_2,\cdots, \p_n$ to expand each image around the
time-averaged image $\langle \mcal I \rangle$:
$\mcal I(t)= \langle \mcal I \rangle+ \sum_{i=1}^n c_i(t) \p_i$. The dynamics of the projection
coefficients are on average governed by the drift field $\mathbf{\Phi}(\cvec)$
and diffusion tensor $\mathbf{D}(\cvec)$ (see Eq.~\eqref{eq:Langevin_components}).

In the simulated data of the two-bead model,
the first four principal components satisfy criteria 1) and 2)
introduced in Sec. ~\ref{sec:principle} (Fig.~\ref{Fig2}c). Interestingly, $\p_1$ and $\p_2$
resemble the in-phase and out-of-phase motion of the two beads,
respectively and should suffice to reproduce the dynamics of $(x_1(t),x_2(t))$, consistently with our third truncation criterion. The components $\p_3$ and $\p_4$ appear to mostly represent the isolated
fluctuations of the hot and cold beads and mainly account for the nonlinear details of the image representation. The first four components, however, allow for an adequate reconstruction of the
original images (Fig.~\ref{Fig2}d).

From the recorded trajectories in $\p_1 \times \p_2$ space we can already
infer key features of the system's dynamics using  SFI. Specifically, we infer the force and diffusion fields (Fig.~\ref{Fig2}e). In the
phase space spanned by the first two principal components, we identify
a stable fixed point at $(0,0)$ (Fig.~\ref{Fig2}e). As may be expected in this case, the $\p_1$-direction (in-phase motion) is less stiff than
the $\p_2$ direction (out-of-phase motion).

The temperature difference between the two beads results in
phase-space circulation, as revealed by the inferred mean velocity field
(Fig.~\ref{Fig2}f). To quantitatively assess the irreversibility
associated with the presence of such phase space currents, we estimate
the entropy production rate of the system $\widehat{\dot{S}}$, which
converges for long enough measurement time
(Fig.~\ref{Fig2}g-inset). Strikingly, already with two principal
components we find good agreement between the inferred and
the exact entropy production rate, capturing from 
$78 \pm 25\%$ at $T_{\rm c}/T_{\rm h}=0.5$) to $88 \pm 7 \%$
of the entropy production at $T_{\rm c}/T_{\rm h}=0.2$ (Fig.~\ref{Fig2}g). Furthermore, the difference between
the exact and inferred entropy production is consistent with the
typical inference error predicted by SFI. As expected, the estimate of the entropy production rate increases with the number of included components. Note that including more modes than the dimension of the physical phase space (in this case 2) can lead to an overestimate of $\dot{S}$ (Fig.~\ref{Fig2}g). Finally, we note that the functional dependence of $\dot{S}$ on $T_{\rm c}/T_{\rm h}$ is fully recovered and, importantly, no significant entropy production is inferred when the bead temperatures are identical (equilibrium).

We can also use the information contained in the first four
principal components to quantitatively infer forces in image-space via
the relation $\Fhat(\I(t))= \sum_{i=1}^4 \widehat{\Fb}_i(\cvec(t)) \p_
i$. Note that while two modes were sufficient to infer $\widehat{\dot{S}}$, 
more modes are needed to reconstruct the full images and
image-force fields as a linear combination of modes.
Importantly, when inferring forces we always subtract from the drift the
spurious force  $\nabla \cdot \mathbf{D}(\cvec)$ arising in overdamped It\^o
stochastic differential equations with multiplicative noise. For
comparison purposes, the exact image force field is obtained directly
from the simulated data as:
$\Fex(t)=[\bar{\I}(\x(t)+\Fb(\x) \Delta t)-\bar{\I}(\x(t))]/\Delta t$. Remarkably, we find good qualitative
agreement between inferred and exact image force fields for specific
realizations of the system, as shown in the kymographs in
Fig.~\ref{Fig2}i (see also Supplementary Movies 2 and 3). Moreover, we find a strong correlation (Pearson
correlation coefficient $\rho=0.93$) between inferred and exact
image-forces. To further quantify the performance of force inference, we compute the relative squared error on the inferred image force field ($\sigmaF=\sum_t \lVert \Fhat(t)-\Fex(t) \rVert^2/\sum_t \lVert \Fhat(t)\rVert ^2 $), which in this case is modest $\sigmaF=0.14$ (Fig. \ref{Fig2}h). 

Thus, with sufficient information, we can use our approach to accurately predict at any instant of time the physical force fields in image space from the Brownian movie, even if the system is out of equilibrium.
Moreover, the results for this simple two-bead system demonstrate the validity of our approach: we reliably infer the non-equilibrium dynamics of this system. Arguably, direct tracking of the two beads is, in this case, a more straightforward approach. However, this changes when considering more general soft assemblies comprised of many degrees of freedom. 

\section{Dissipative Component Analysis: A principled approach to construct the most dissipative components}
\begin{figure*}[h!tb] 
\centering
\includegraphics[width=\textwidth]{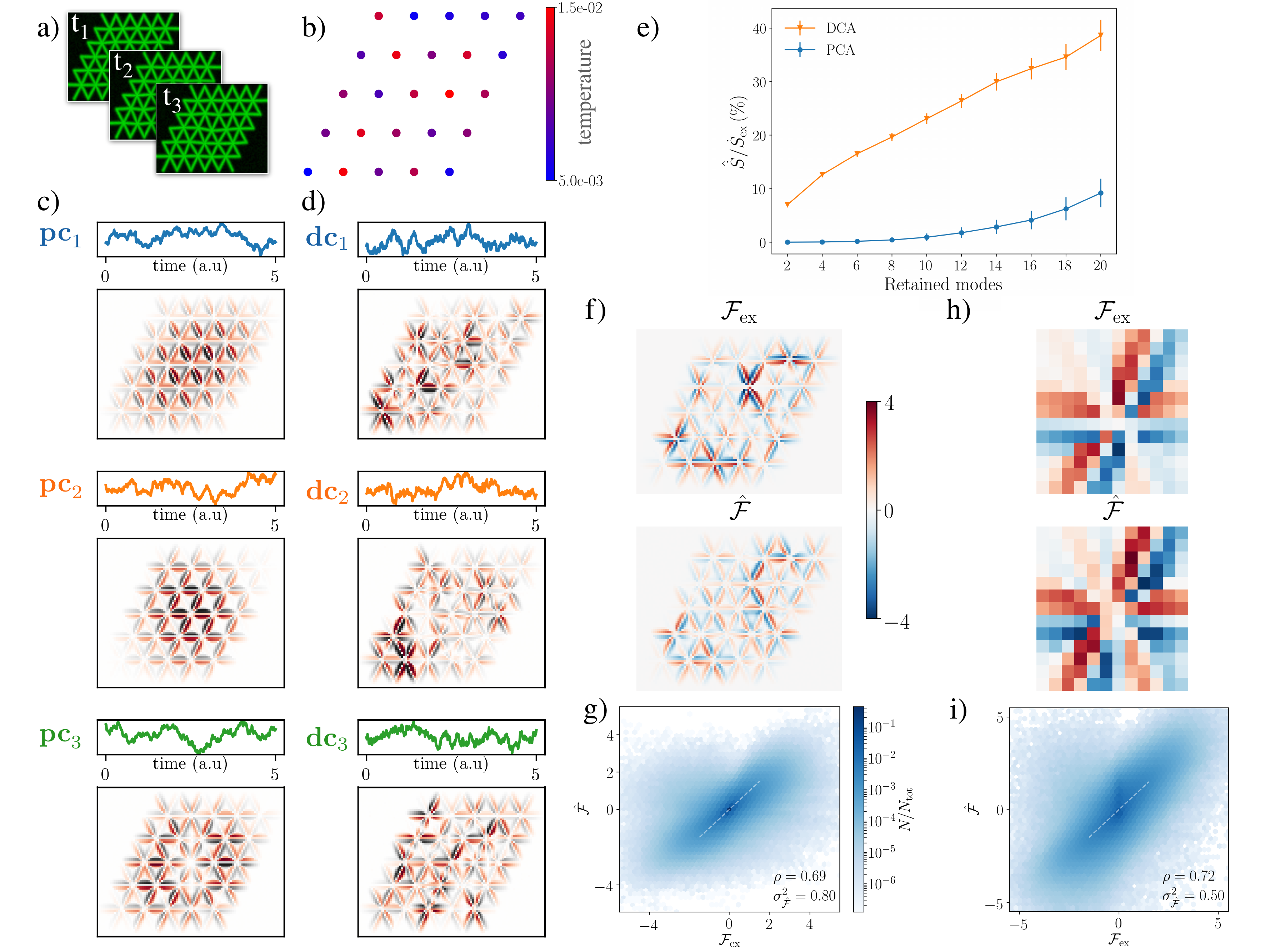}
\caption{\textbf{Learning the non-equilibrium dynamics of Brownian movies of simulated filamentous networks} a) $100 \times 80$ frames of a $5\times5$ filamentous network with fixed boundary conditions driven out of equilibrium by a heterogeneous temperature distribution. b) The  temperatures at the nodes are indicated with a different color ranging from blue (low temperature) to red (high temperature). c-d) Trajectory of the projection coefficient (top) and image-component (bottom) for PCA (c) and DCA (d). e) The estimated entropy production rate $\widehat{\dot{S}}$ as a function of the number of components included in the analysis. See Supplementary Sec. IV for additional data at equilibrium and convergence of the estimates. f) Full network: Comparison of the exact image-force $\mathcal{F}_{\rm ex}$ (top) to the inferred one $\hat{\mathcal{F}}$ (bottom) at a selected instant of time with $50$ PCs. g) Scatter plot of the exact force field $\mathcal{F}_{\rm ex}$  vs. the inferred force field $\hat{\mathcal{F}}$ for different pixels and time points with $50$ PCs (data has been binned for visualization purposes). At the bottom right the Pearson correlation coefficient $\rho$ and the relative error squared $\sigmaF$ are indicated. h) Single cropped patch: Comparison of the exact image-force $\mathcal{F}_{\rm ex}$ (top) to the inferred one $\hat{\mathcal{F}}$ (bottom) at a selected instant of time with $20$ PCs. Arrows indicate the deterministic velocity field. Colorbar same as in f). i) Piecewise reconstruction of force inference for full network from cropped patches: Scatter plot of the exact force field $\mathcal{F}_{\rm ex}$  vs. the inferred force field $\hat{\mathcal{F}}$ for different pixels and time points with $20$ PCs. Colorbar same as in g). All results have been obtained with a trajectory of $10^6$ time steps, $\Delta t=0.005$ and $100 \times 80$-pixels frames for the full network. The SFI routine was employed with a first order polynomial basis for the inference of forces and diffusion fields, and noise-corrected diffusion estimates.}
\label{Fig3}
\end{figure*}

To expand the scope of our approach, we next consider  a more complex scenario inspired by cytoskeletal assemblies: a  network of elastic filaments (Fig.~\ref{Fig3}a). The filaments are modeled as Hookean springs that connect two neighboring nodes on a triangular network. The  Langevin equation for the two-dimensional displacement $\x_i$ of the network's $i$-th node is given by \Eq{eq:Lang_initial}. In this case, the force acting on node $i$ is $\Fb_i(\x)=-\sum_{j \sim i} \frac{k}{\gamma}(\lVert \x_{i,j}(t) \rVert-\ell_0)\hat{\x}_{i,j}$, $\mathbf{x}_{i,j}=\mathbf{x}_{i}-\mathbf{x}_{j}$, $\hat{\mathbf{x}}_{i,j}$.
is the corresponding unit vector, and the sum runs over nearest-neighbors $j$ of $i$. Similarly to the two-bead model (Fig.~\ref{Fig2}), we drive the system out of equilibrium by imposing spatially heterogeneous node temperatures drawn randomly from a uniform distribution, as shown in Fig. \ref{Fig3}b. We impose rigid boundary conditions to avoid rotations and diffusion of the system as a whole.

We simulate the dynamics of a $5 \times 5$ network: for each time step we create an image in which neighboring nodes are connected by filament segments  and measurement noise is added to generate a  Brownian movie (see Supplementary Material Sec. II, Fig.~\ref{Fig3}a, and Supplementary Movie 4). In this spatially extended system, generated \ from an underlying dynamics with  $50$ degrees of freedom, it is not obvious based on the recorded Brownian movie how to select and analyze the relevant degrees of freedom.

We start our movie-based analysis by employing PCA to reduce the dimensionality of the image data  (Fig.~\ref{Fig3}c). For this set of simulation data, our truncation criteria indicate that the maximum number of retainable components is roughly $50$,  consistent with the number of degrees of freedom in the underlying dynamics. Although we greatly reduced dimensionality of the image data using this truncation, it is still intractable to infer dynamics in a $50$-dimensional space due to limited statistics. However, even a subset of these modes may suffice to glean useful information about the system's non-equilibrium dynamics. Therefore, as a first attempt, we identify the modes that retain most of the variance via PCA and infer the dynamics in increasingly larger PC-space via SFI. This allows us to infer the retained percentage of entropy production rate as a function of the number of principal components considered (Fig.~\ref{Fig3}e). In contrast to the two-beads case, we observe that in this more realistic scenario  we recover less than $10 \%$ of the system's entropy production rate with the first twenty PCs. Indeed, PCA is designed to find modes that capture the most variance in the image data, and large variance, does not necessarily imply large dissipation. Thus, in this case, PCA fails at selecting components that capture a substantial fraction of the entropy production rate. 

Our goal is to infer the system's non-equilibrium dynamics. We thus propose an alternative way of reducing data dimensionality that spotlights the time-irreversal contributions to the dynamics, which we term Dissipative Component Analysis (DCA). DCA represents a principled approach to determine the most dissipative pairs of modes for a linear system with state-independent noise (see Supplementary Material Sec. III). For such a linear system, there exists a set of component pairs for which the entropy production rate can be expressed as a sum of independent positive-definite contributions, which can be ranked by magnitude. After a suitable truncation, this basis ensures that the components with the largest entropy production rate are selected. While the approach is only rigorous for a linear system with state-independent noise, we demonstrate below that this method  also performs well for more general scenarios.

DCA relies on the measurement of an intuitive trajectory-based non-equilibrium quantity: the area enclosing rate (AER) matrix $\mathcal{A}$ associated to a general set of coordinates $\y$. The elements of the AER matrix, in It\^o convention, are defined by \cite{ghanta_fluctuation_2017,gonzalez_experimental_2019,gradziuk_scaling_2019,frishman_learning_2019}
\begin{equation}
\label{eq:AER}
\mathcal{A}_{ij}=\frac{1}{2} \langle y_j \dot{y}_i- y_i\dot{y}_j \rangle,
\end{equation} 
where $y_i$ denotes the $i$-th coordinate centered around its mean value and $\langle \cdot \rangle$ a time average.
This non-equilibrium measure quantifies the average area enclosed by the  trajectory  in phase space per unit time. Importantly, the AER is tightly linked to the entropy production rate. Specifically, for a linear system $\dot{S}~=~\Tr (\mathcal{A} C^{-1} \mathcal{A}^T D^{-1})$ where the covariance matrix $C_{ij} = \langle y_i y_j \rangle $. DCA identifies a basis of vector pairs
$\{\mathbf{dc}_1,\mathbf{dc}_2; \mathbf{dc}_3,\mathbf{dc}_4;\ldots\}$
 that simultaneously transforms $C$ to the identity and diagonalizes $\mathcal{A}\mathcal{A}^{T}$ (see Supplement Sec. III). By doing so, DCA naturally separates the entropy production rate into independent contributions that can be readily ordered by magnitude, \emph{i.e.} $\dot{S}~=~\dot{S}_{\mathbf{dc}_1,\mathbf{dc}_2} + \dot{S}_{\mathbf{dc}_3,\mathbf{dc}_4} + \cdots$ with $\dot{S}_{\mathbf{dc}_1,\mathbf{dc}_2} > \dot{S}_{\mathbf{dc}_3,\mathbf{dc}_4} > ~\cdots~$. Truncating the basis of dissipative components using the aforementioned criteria, allows us to identify a few components that are assured to maximally contribute to the dissipation of the system.

To test the performance of DCA, we revisit the network simulations. We first perform PCA to reduce noise and dimensionality. Subsequently, we perform DCA with these first $50$ principal component coefficients as input.  The dissipative components are very different from the principal components (Fig.~\ref{Fig3}d): while the principal components seem to capture the collective displacement modes of the filaments, the dissipative components appear to reflect the local temperature inhomogeneities in the network. Strikingly, DCA allows us to recover a substantial portion of the total entropy production rate (almost $40 \%$ with $20$ components) performing about twenty times better than the PCA-based approach, as shown in Fig.~\ref{Fig3}e. 

Even when we recover only a fraction of the entropy production, our inference approach yields additional insightful information about the dynamics in the system, such as force field estimates. To investigate to what extent our movie-based learning approach reconstructs the elastic forces exerted by the network's filaments, we compare the inferred force field in image space to the exact one. For this purpose, we employ PCA in our dimensional reduction scheme, which can be used both in and out of equilibrium. Remarkably, even in this large network we find that the inferred force field in image space can capture the basic features of the exact force field, as shown in Fig. \ref{Fig3}f-g and in Supplementary Movies 5 and 6. However, inferring image force fields with high accuracy for the full $5 \times 5$ network is challenging due to the curse of dimensionality~\cite{bishop_pattern_2006}, as confirmed by the sizeable force inference error reported in Fig. \ref{Fig3}g. 

To perform accurate force inference on large systems, we perform a piecewise learning of spatially cropped Brownian movies. 
Put simply, we can exploit the locality of the interactions in the system to extract information about local forces from local  dynamics in image space. More specifically, we divide each frame of the movie into disjoint cropped patches and reconstruct image-forces in each patch separately, as shown in Fig. \ref{Fig3}h (see Supplementary Material Sec. VI). Then, we can use the force field inferred in each patch to reconstruct the force field for the full image and thus for the full network. This procedure not only improves force inference, as shown in Fig. \ref{Fig3}i, but also has the advantage of being scalable: While image force inference for a whole system becomes unfeasible for large assemblies, the cropping procedure can yield accurate force estimates independent of system size.

In sum, we have demonstrated how a combination of PCA and DCA allows us to recover a substantial fraction of the entropy production in a complex scenario such as a $5 \times 5$  network with measurement noise. Note the dynamics in image space in this system is described by a non-linear Langevin equation with multiplicative noise. Despite this complexity, our scalable approach is  able to infer the basic features of the  force field.

\section{Discussion}
\noindent We considered the dynamics of movies of time-lapse microscopy data. Under the assumptions outlined in Sec.~\ref{sec:principle}, these movies undergo Brownian dynamics in image space: the image-field obeys an overdamped Langevin equation of the form of Eq.~\eqref{eq:Langevin_images_new}.  Rather than tracking  selected degrees of freedom, we propose to analyze the Brownian movie as a whole. 

Our approach is based on constructing a reduced set of relevant degrees of freedom to reduce dimensionality, by combining PCA with a new method that we term Dissipative Component Analysis (DCA). In the limit of a linear system with state-independent noise, DCA provides a principled way of constructing and ranking independent dissipative modes. 
The order at which we truncate is an important trade-off parameter of this method: on the one hand we wish to
  significantly reduce the dimensionality of the data, on the other
  hand we need to include enough components to retain the information
  necessary to infer the system's dynamics. After the dimensional reduction, we infer the stochastic dynamics of the system, revealing the force field, phase space currents, and the entropy production rate in this basis. This information can then be mapped back to image-space to provide estimators for the stochastic dynamics of the Brownian movie. We illustrated our approach on simulated data of a minimal two-beads model and on filamentous networks in both equilibrium and non-equilibrium settings, and showed that it is robust in the presence of  measurement noise. Beyond providing controlled lower bounds of the entropy production rates directly from the Brownian movie, our approach yields estimates of the force-fields in image space for an instantaneous snapshot of the system and we demonstrated that this approach can be scaled up to large systems. Thus, we provide in principle an alternative to microscopic force and stress sensing \mbox{methods}~\cite{sawada_force_2006,grashoff_measuring_2010,lucio_chapter_2015,han_cell_2018}.
  
We  focused here on a class of soft matter systems termed ``active viscoelastic solids"~\cite{fletcher_active_2009,gnesotto_broken_2018}. Such systems include active biological materials such as cytoskeletal assemblies ~\cite{mizuno_nonequilibrium_2007,jensen_mechanics_2015,koenderink_active_2009,brangwynne_nonequilibrium_2008}, membranes~\cite{turlier_equilibrium_2016,betz_atp-dependent_2009,ben-isaac_effective_2011}, chromosomes ~\cite{weber_nonthermal_2012}, protein droplets~\cite{brangwynne_active_2011}, as well as active turbulent solids ~\cite{hemingway_active_2015} and colloidal systems~\cite{aranson_active_2013}. Although these structures are constantly fluctuating both due to energy-consuming processes (\emph{e.g.} rapid contractions generated by molecular motors) and thermal motion, they do not exhibit macroscopic flow. 
Useful insights into the properties of such systems have been obtained via different non-invasive techniques. Typically, these techniques employ time traces of tracked object to extract
information about the active processes governing the non-equilibrium behavior~\cite{mura_nonequilibrium_2018, gnesotto_nonequilibrium_2019, gladrow_broken_2016, battle_broken_2016, seara_entropy_2018,turlier_equilibrium_2016,betz_atp-dependent_2009}. Often, however, it is not \textit{a priori} obvious which physical degrees of freedom should be tracked, how tracking can be performed in fragile environments, and to what extent the dynamical information about the system of interest is encoded in the measured trajectories~\cite{seara_dissipative_2019}. While tracking-free approaches have been proposed to obtain rheological information of a system under equilibrium conditions~\cite{edera_differential_2017}, our approach offers an alternative to tracking that can provide information on dissipative modes and the instantaneous force fields of a fluctuating non-equilibrium system. 

In summary, we presented a viable alternative to traditional analysis techniques of high-resolution video-microscopy of soft living assemblies. Indeed, we envision experimental scenarios where our approach may serve as a guide, providing novel insights by disentangling the deterministic and stochastic components of the dynamics, and by helping to identify the source of thermal and active forces as well as the dissipation in the system. Overall, our movie-based approach constitutes an adaptable tool that paves the road for a systematic, non-invasive and tracking-free analysis of time-lapse data of soft and living systems.

\section*{Acknowledgements}
We thank C. Schmidt, F. Mura, S. Ceolin and I. Graf for many stimulating discussions. This work was Funded by the Deutsche Forschungsgemeinschaft (DFG, German Research Foundation) under Germany's Excellence Strategy – EXC-2094 – 390783311 and by the DFG grant 418389167.

\bibliography{References_Brownian_Movies}



\section*{Supplementary material}

# Learning the Non-Equilibrium Dynamics of Brownian Movies: Supplementary Material

## I. NUMERICALLY INTEGRATING THE BROWNIAN DYNAMICS

We simulate the stochastic dynamics of the two-beads model by numerically integrating the overdamped Langevin equation for the beads' displacements $\mathbf{x} = (x_1, x_2)$ (Eq. 1 main text), with $\mathbf{F}(\mathbf{x}) = \mathbf{A}\mathbf{x}$, $A_{ij} = (1 - 3\delta_{ij})k/\gamma$, $D_{ij} = \delta_{ij}k_B T_i/\gamma$. We discretize the equation of motion for the two beads using an Euler scheme with discretization step $\Delta t$. Thus, the discretized equation of motion after $n$ time steps for the $i$-th bead reads: $x_i((n+1)\Delta t) = x_i(n\Delta t) + \sum_{j=1,2} A_{ij} x_j(n\Delta t)\Delta t + \sum_{j=1,2} \sqrt{2D_{ij}\Delta t}\xi_j$, where $\xi_j$ is a random number drawn from a normal distribution with mean zero and average one. We initialize the simulation with the beads in their rest state and we only record the positions of the beads after an equilibration time $t_{\text{eq}} = 10^5 \Delta t$ to allow the dynamics to reach steady state. The parameters for the results of Fig. 2 of the main text are: $\Delta t = 0.01$, $k = 2$, $\gamma = 1$, $k_B = 1$, $T_1 = 1$ and $0.2 < T_2 \le 1$.

The dynamics of the $5 \times 5$ spring network is generated in a similar way. In this case, we discretize the overdamped Langevin equation for the nodes' positions $\mathbf{x}$ with time step $\Delta t = 0.005$. For the network the elastic force acting on node $i$ reads: $F_i(\mathbf{x}) = -\sum_{i \sim j} \frac{k}{\gamma}(\|\mathbf{x}_{i,j}(t)\| - \ell_0)\hat{\mathbf{x}}_{i,j}$, $\mathbf{x}_{i,j} = \mathbf{x}_i - \mathbf{x}_j$, $\mathbf{x}_{i,j}$ is the unit vector between nodes $i$ and $j$, $k = 4$, and $\ell_0 = \gamma = k_B = 1$. The heterogeneous temperatures at the different nodes are chosen randomly from 25 uniformly spaced values in the interval $[T_0 - T_0/2, T_0 + T_0/2]$ with $T_0 = 10^{-2}$. The simulation is initialized with the network in its rest state and we wait an equilibration time $t_{\text{eq}} = 10^5 \Delta t$ before recording trajectories.

## II. GENERATING THE BROWNIAN MOVIES

We first outline the procedure to generate a Brownian movie for the two-beads model (see Fig. 2 of the main text). The input consists of the numerically generated position trajectories of the two beads. We then transform the trajectories from position space to image space into pixel units (we used a $40 \times 20$ pixel grid). Specifically, we set the image pixel intensities at a given time point by centering a radially symmetric Gaussian function centered at the bead's position, with amplitude 1 and variance 9 pixels . Finally, to simulate measurement noise in a simple way, we add uncorrelated white noise sampled uniformly from $[0, a]$ ($a = 0.1$, i.e. in Fig. 2 of the main text) independently at each pixel. As in real imaging devices, pixels are saturated at intensity 1, thus any intensity larger than 1 is truncated to one.

Next, we briefly explain how we generate a movie for the $5 \times 5$ network. The $N \times 50$-dimensional position array which is the output of the numerical integration of the Langevin equation is transferred to a custom Python routine that directly plots all lines connecting neighboring nodes at each time step onto a $100 \times 80$ grid. Specifically, the pixel intensities decay with the distance from each line as a Gaussian function with amplitude 0.8 and variance 2 pixels. Finally, to simulate measurement noise, we add uncorrelated white noise sampled uniformly from $[0, a]$ ($a = 0.08$–10% of the maximum intensity– in Fig. 3 of the main text).

## III. INFERRING THE DISSIPATIVE MODES: DISSIPATIVE COMPONENT ANALYSIS

The aim of Dissipative Component Analysis (DCA) is to infer a set of modes that maximize dissipation or, more precisely, the entropy production rate. This method is a principled approach only for a linear dynamical system with constant diffusion. However, as we demonstrate in the main text, this method can be successfully employed in high-dimensional situations when dealing with image-data, when the dynamics is close to linear (close to the stable fixed points of the system). In such cases DCA can reduce the dimensionality by exploiting the non-equilibrium character of the system, as outlined below.

We consider a generic linear system described by an $n$-dimensional column-vector of coordinates $\mathbf{y}$ that obeys the Langevin equation

$$\frac{d\mathbf{y}}{dt}(t) = \mathbf{A}\mathbf{y}(t) + \sqrt{2D}\boldsymbol{\xi}(t)\,, \tag{S1}$$

where $\mathbf{A}$ is the interaction matrix and $\mathbf{D}$ the diffusion matrix. Note that $\mathbf{D}$ and $\mathbf{A}$ may in general not satisfy detailed balance and the system may thus be out of equilibrium.

As a preliminary step we perform principal component analysis (PCA) on data obtained simulating the time-evolution described by Eq. S1 for $N$ time-steps: we first compute the covariance matrix $\mathbf{C} = \frac{1}{N}\sum_{t=1}^{N}(\mathbf{y}(t)-\langle\mathbf{y}\rangle)\cdot(\mathbf{y}^T(t)-\langle\mathbf{y}\rangle^T)$, where $\langle\mathbf{y}\rangle = \frac{1}{N}\sum_{t=1}^{N}\mathbf{y}(t)$. We then retain the first $m < n$ eigenvectors of $\mathbf{C}$ (see Sec. V for details on the truncation criteria), ordered by magnitude of the associated eigenvalues, and use them to construct the $m \times n$ transformation matrix $\mathbf{E}$. The time evolution of the system projected onto the PC-coordinates is then $\mathbf{y}_{pca}(t) = \mathbf{E}^T\mathbf{y}(t)$. In this basis, the covariance matrix $\mathbf{C}_{\text{pca}}$ is diagonal with the ordered eigenvalues as diagonal entries. This preliminary step is useful for two reasons: it reduces dimensionality and it conveniently filters out measurement noise from the images. Next, we transform the data into covariance identity coordinates (cic), in which the covariance matrix is the identity. This is accomplished by $\mathbf{y}_{\text{cic}}(t) = \mathbf{C}_{\text{pca}}^{-1/2}\mathbf{E}^T\mathbf{y}(t)$.

In the next step, we focus on the non-equilibrium character of the system and compute the area-enclosing-rate matrix (AER) $\dot{\mathcal{A}}$ in CIC coordinates [1–3]:

$$\mathcal{A}_{\text{cic},ij} = \frac{1}{2t_{\text{tot}}}\sum_{t=1}^{N}[y_{\text{cic},i}(t)\Delta y_{\text{cic},j}(t) - y_{\text{cic},j}(t)\Delta y_{\text{cic},i}(t)], \tag{S2}$$

where $t_{\text{tot}} = N\Delta t$ is the total simulation time and $\Delta y_i$ denotes the displacement of the $i$-th coordinate between two successive time-steps. Each element $\mathcal{A}_{ij}$ of the AER matrix corresponds to the area that the trajectory encloses on average in the plane $(y_i, y_j)$ per unit time. This area enclosing rate quantifies broken detailed balance in the system and is zero in thermal equilibrium. Having defined the AER allows us to conveniently write the total entropy production of the system as [2, 4]:

$$\dot{S} = \text{Tr}(\mathcal{A}_{\text{cic}}\mathcal{A}_{\text{cic}}^T\mathbf{D}_{\text{cic}}^{-1}), \tag{S3}$$

where $\mathbf{D}_{\text{cic}} := \frac{1}{2t_{\text{tot}}}\sum_t \Delta\mathbf{y}_{\text{cic}}(t)\Delta\mathbf{y}_{\text{cic}}^T(t)$. It is now key to observe that the matrix product $\mathcal{A}_{\text{cic}}\mathcal{A}_{\text{cic}}^T$, appearing in the expression for the entropy production rate Eq. S3, is real and symmetric and thus admits a real orthonormal basis of eigenvectors. Moreover, since $\mathcal{A}_{\text{cic}}$ is antisymmetric, all non-zero eigenvalues of $\mathcal{A}_{\text{cic}}\mathcal{A}_{\text{cic}}^T$ are two-fold degenerate. Furthermore, note that the orthonormal basis of $\mathcal{A}_{\text{cic}}\mathcal{A}_{\text{cic}}^T$ is unique up to rotations in the two-dimensional eigenspaces that correspond to the same eigenvalue. Importantly, in these special covariance identity coordinates (scic), the total entropy production rate reads

$$\dot{S} = \sum_{i \in odd} \lambda_i[(\mathbf{D}_{\text{scic}}^{-1})_{ii} + (\mathbf{D}_{\text{scic}}^{-1})_{i+1\,i+1}], \tag{S4}$$

with $\lambda_i$ being the eigenvalues of $\mathcal{A}_{\text{cic}}\dot{\mathcal{A}}_{\text{cic}}^T$. We refer to the corresponding eigenvectors as the dissipative components.

## IV. DEPENDENCE OF ENTROPY PRODUCTION RATES ON THE TRAJECTORY LENGTH

The entropy production rate is a semi-positive definite quantity: at steady state $\dot{S} \geq 0$. Given finite-length data, the estimate of the entropy production rate will be biased. While this bias can be computed analytically for homogeneous diffusion coefficients [5], this may be difficult for space-dependent diffusion coefficients and in the presence of measurement noise. Given that we are here concerned with finite-size data of systems with multiplicative noise partially corrupted by measurement noise, we use the following approach to reduce the bias of the entropy production rate and, correspondingly, to avoid overfitting: We separate our data set of length $N$ into two independent and successive sets, a training set of length $m$ and a test set of length $n = N - m$. The results in Fig. 3 of the main text are obtained with $m = N/10$. We first infer relevant components using the training set, and we then project the test set onto these components and infer the corresponding entropy production rate, as shown in Supplementary Fig. 1a . Although entropy production rate estimates remain weakly positively biased for short trajectories, the bias approaches zero for long trajectories, as shown in Supplementary Fig. 1b for the spring network with uniform temperatures (equilibrium). Note, however that our error bar estimates always intersect zero for all trajectory lengths (Supplementary Fig. 1b). When the network is out of equilibrium, the entropy production rate estimates converge to non-zero values for long trajectories, as shown in Supplementary Fig. 1c.

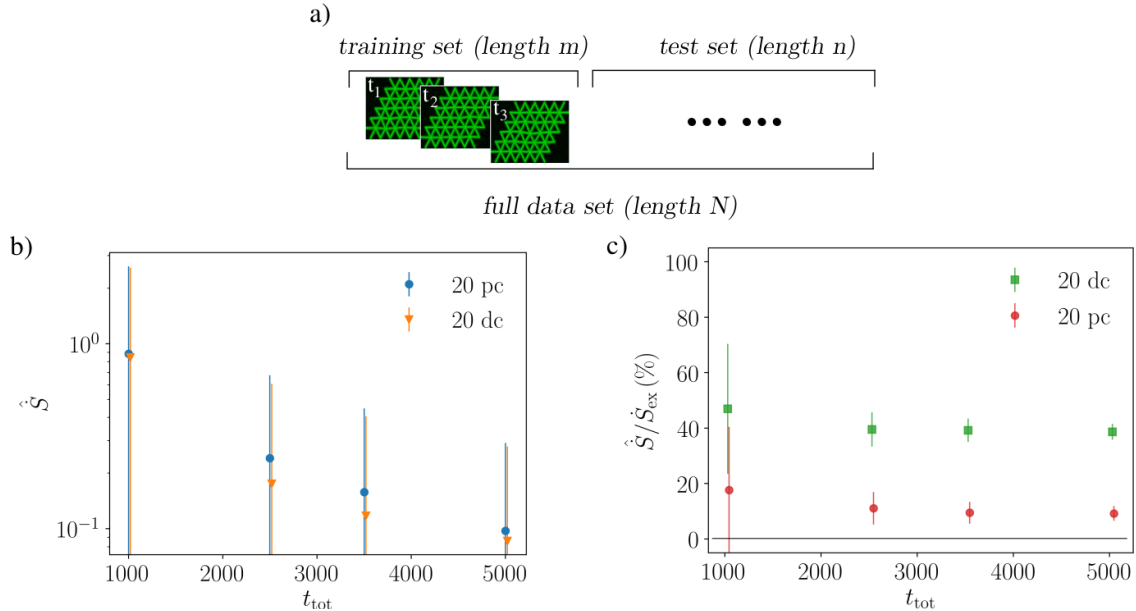

Supplementary Figure 1. a): Schematic of the training/test set splitting procedure: the full trajectory (length $N$) is split into a training set (length $m$) and into a test set (length $n$). b) Decay to zero of the entropy production rate bias (estimated with 20 pc-blue dots and 20 pc-orange triangles) as a function of the trajectory length at equilibrium. c) Convergence of the entropy production rate (estimated with 20 pc-red dots and 20dc-green squares) as a function of the trajectory length. The parameters of the simulations and noise level are the same as in Fig. 3 of the main text. Equilibrium is obtained by setting all temperatures equal to $T_0 = 10^{-2}$.

## V. DIMENSIONAL REDUCTION: TRUNCATION CRITERIA

For the Brownian-movie learning procedure it is important to reduce the dimensionality of image data to a more tractable number of components. Therefore, we require criteria to decide on the maximum number of components that we consider in our analysis of the stochastic dynamics. Two main limiting effects arise due to the finite length of trajectories and measurement noise.

### 1) Noise floor

We start by asking what is the maximum number of components that we can distinguish from a noise floor set by the imaging noise and the finite length of the data. Our image data is a matrix $\mathbf{X}$ of $t_{\text{tot}}$ (total simulation time) rows and $L \times W$ (total number of pixels in a single image) columns. We first estimate the principal components — the normalized eigenvectors of the covariance matrix of image data — and sort these components according to the magnitude of the corresponding eigenvalues. To determine the noise floor, we eliminate temporal correlations in the image data by shuffling the values of $\mathbf{X}$ separately along each of its columns [6]. What we obtain is a shuffled data set $\mathbf{X}_{\text{shuffled}}$ for which we can also compute principal components and eigenvalues. The largest eigenvalue of the covariance matrix of $\mathbf{X}_{\text{shuffled}}$ yields the noise floor. Thus, we truncate the basis of principled components to exclude components with eigenvalues below this noise floor. To illustrate this procedure, a plot of the eigenvalues for $\mathbf{X}$ together with the noise threshold is shown in Supplementary Fig. 2 a-b for the two beads model and for the filamentous network.

### 2) Resolution of the dynamics

Criterion 1) ensures that the components are distinguishable from imaging noise, which is a static property of the data. The Brownian-movie analysis is concerned with the dynamics. We thus want to make sure that we can resolve the dynamics of the components selected with criterion 1). This is a necessary condition to

infer force and diffusion fields in image-space. A criterion for selecting components whose dynamics can be resolved using SFI is based on computing the autocorrelation function of the projection coefficients (**c** in the main text) centered around their average value ($c_i(t) \to c_i(t) - \langle c_i \rangle$):

$$C_i(n\Delta t) = \frac{\sum_{t=1}^{N-n\Delta t} c_i(t + n\Delta t) c_i(t)}{\sum_{t=1}^{N} c_i^2(t)} \,. \tag{S5}$$

We are only able to resolve the dynamics if $c_i(t)$ does not decorrelate too fast, i.e. if $C_i(n\Delta t)$ does not decay to zero in a time comparable to the time-step $\Delta t$. We therefore employ the following criterion: we only retain components for which $|C_i(\Delta t) - C_i(0)| < 0.25$. We applied criterion 2) to the two-beads data and to the network data and plot the results in Supplementary Fig. 2 c-d. Criterion 2) is clearly sensitive both to the time resolution $\Delta t$ and to the signal to noise ratio in the trajectories.

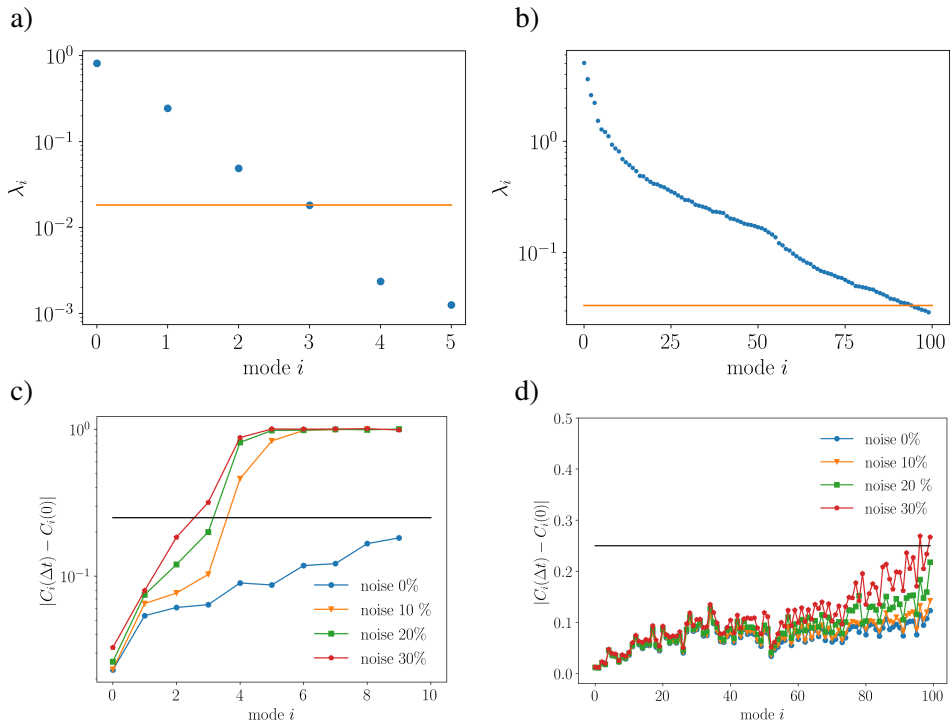

Supplementary Figure 2. a-b): Eigenvalues $\lambda_i$ of the covariance matrix for the data **X** (blue markers) together with the noise floor (largest eigenvalue of $\mathbf{X}_{\text{shuffled}}$-orange line) for the two-beads model (a) and the filamentous network (b). In panels (a) and (b) the noise level on the image is 10%. c-d): Decrease (absolute value) of the autocorrelation function of principal component coefficients after one time-step at different noise levels for the two-bead model (c) and the filamentous network (d). The solid line indicates the 25% level used in our criterion. Panel a-c (Panel b-d): same simulation parameters as Fig. 2 (Fig. 3) of the main text.

### 3) Dimension of the physical phase space

For a physical system $\mathbf{x}(t)$ with $d$ observable degrees of freedom, the ideal images of the system $\bar{\mathcal{I}}(t)$ form a $d$-dimensional manifold in the $(L \times W)$-dimensional image space. The registered images $\mathcal{I}(t)$, which include the measurement noise, lie in a neighborhood of the manifold. For this reason it may be possible to project the image-trajectory on a $d$-dimensional linear subspace, without loosing information about the original dynamics of $\mathbf{x}(t)$. To determine this linear subspace we perform PCA on the whole image set and order the principal modes according to their variance. Because of the curvature of the manifold such analysis may not reveal the physical phase space dimension $d$.

To resolve this problem we randomly choose an image $\mathcal{I}(t_0)$ and look for all the points of the image trajectory lying in a sphere of radius $r$ around it. Having found the set $B_r[\mathcal{I}(t_0)] = \{\mathcal{I}(t) : \|\mathcal{I}(t_0) - \mathcal{I}(t)\|^2 \leq r^2\}$ we proceed to perform a local version of PCA on the sets $B_r[\mathcal{I}(t_0)]$. As we decrease the radius $r$, we begin to probe the manifold in a region where it is approximately flat. Consequently, we observe a gap appearing in the plot of the eigenvalues of the covariance matrix (see Supplementary Fig. 3), indicating the actual dimension of the manifold.

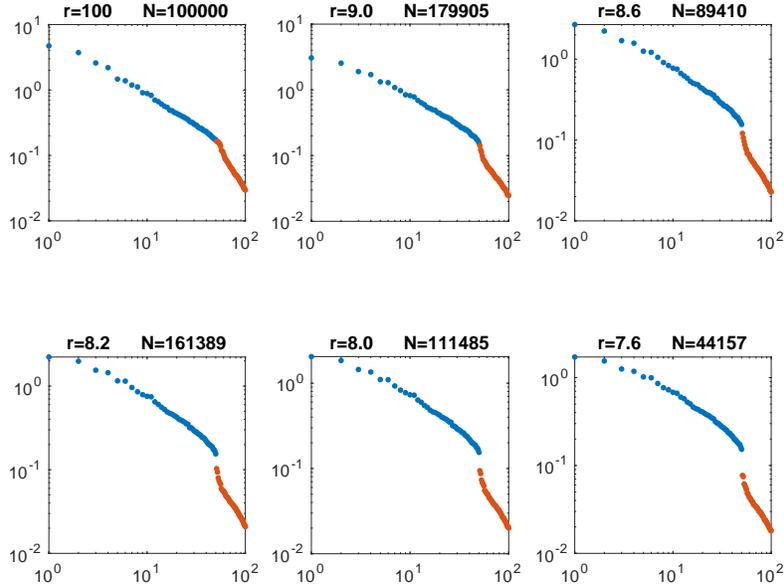

Supplementary Figure 3. Plots of the variances of the first 100 local PCA modes calculated for $N$ images inside a sphere $B_r[\mathcal{I}(t_0)]$. The images used here represent the dynamics of the $5 \times 5$ network (50 degrees of freedom). The variances of the first 50 principal components are plotted in blue, the remaining ones in red. The distribution of the variances of the modes changes as we decrease the radius of the sphere $r$. For small radii a gap appears at the 50th mode.

## VI. PATCHING PROCEDURE FOR FORCE INFERENCE

In this section, we briefly outline the patching procedure that allowed us to improve force inference for the network and make the approach scalable to large systems. First, we tessellate every frame of our movie into 25 disjoint patches, as shown in white in Supplementary Fig. 4. To infer forces inside a patch (for example the red patch in Supplementary Fig. 4), we learn the dynamics of a slightly larger region (indicated in blue in Supplementary Fig. 4. This region approximately encloses the parts of the image that interacts with those inside the patch. We then compare the inferred force field in the smaller patch (red in Supplementary Fig. 4). Repeating this procedure for every patch yields the plot of Fig. 3i of the main text.

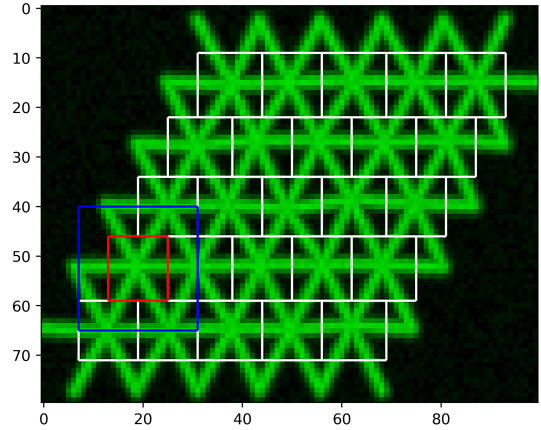

Supplementary Figure 4. Schematic of the patching procedure: The 25 patches in which every frame of the movie is divided are indicated in white. The dynamics is learned inside a larger blue patch and forces inferred in the smaller red patch (see Fig. 3h of the main text for the force pixel map in this patch).


\end{document}